\documentclass[a4paper,fleqn,usenatbib]{mnras}

\usepackage{newtxtext,newtxmath}
\usepackage[T1]{fontenc}
\usepackage{ae,aecompl}
\usepackage{graphicx}
\usepackage{amsmath}
\usepackage{makecell}

\title[Star Formation Reignition in Dwarf Galaxies]{Reignition of Star Formation in Dwarf Galaxies}
\author[A. C. Wright et al.]{
Anna C. Wright,$^{1}$\thanks{E-mail:awright@physics.rutgers.edu}
Alyson M. Brooks,$^{1}$
Daniel R. Weisz$^{2}$ and
\newauthor{ Charlotte R. Christensen$^{3}$}
\\
$^{1}$Department of Physics \& Astronomy, Rutgers, The State University of New Jersey, 136 Frelinghuysen Road, Piscataway, NJ 08854, USA\\
$^{2}$Department of Astronomy and Theoretical Astrophysics Center, University of California Berkeley, Berkeley, CA 94720, USA\\
$^{3}$Department of Physics, Grinnell College, Noyce Science Center, 1116 Eighth Ave, Grinnell, IA 50112, USA
}
\pubyear{2018}
\begin{document}
\label{firstpage}
\pagerange{\pageref{firstpage}--\pageref{lastpage}}
\maketitle
\begin{abstract}
The Local Group hosts a number of dwarf galaxies that show evidence of periods of little to no star formation. We use a suite of cosmological simulations to study how star formation is reignited in such galaxies. We focus on isolated galaxies at $z=0$ with halo masses between 9.2$\times$10$^8$ M$_\odot$ and 8.4$\times$10$^9$ M$_\odot$, where star formation is typically shut off by reionization or by supernova feedback. Nearly 20\% of these simulated galaxies later restart star formation, due to interactions with streams of gas in the intergalactic medium, indicating that this mechanism is relatively common in this mass range and that many isolated dwarfs at $z=0$ may not have been isolated throughout their histories. While high ram pressure interactions lead to stripping, the encounters that reignite star formation are low density and/or low velocity and thus low ram pressure, resulting in compression of the hot gas in the halos of our dwarfs. The gas mass bound up in hot halos can be substantial -- at least an order of magnitude greater than the mass contained in HI. Consequently, we find that dwarfs that have experienced reignition tend to be more HI-rich and have a higher M$_{HI}$/M$_{*}$ ratio at $z=0$ than galaxies with continuous star formation. Using this fact, we identify galaxies in the Local Volume that might have ``gappy" star formation histories, and can be studied by the Hubble Space Telescope or the James Webb Space Telescope.
\end{abstract}

\begin{keywords}
galaxies: dwarf --  galaxies: stellar content -- galaxies: evolution -- Local Group -- intergalactic medium
\end{keywords}

\maketitle
\section{Introduction}
Recent observations suggest that dwarf galaxies within the Local Volume have remarkably diverse star formation histories \cite[e.g.,][]{mateo1998dwarf,orban2008delving,tolstoy2009star,weisz2011acs,weisz2014star,gallart2015acs,skillman2017islands}. Although there is broad agreement within morphological types, there is considerable individual variation from galaxy to galaxy. This diversity in star formation histories (SFHs) has primarily been studied in the context of Milky Way satellites, which have a clearly identifiable source of environmental influence. As they pass through the cloud of gas surrounding their host, many satellites lose their own gas via ram pressure stripping, leading to premature truncation or dampening of star formation \cite[e.g.,][]{kravtsov2004tumultuous,hester2006ram,yozin2015transformation,weisz2015star,fillingham2016under,emerick2016gas,brown2017cold}. Others have been observed to experience bursts of star formation as a result of the same force \cite[e.g.,][]{fujita1999effects,bekki2003starbursts,hester2006ram,kronberger2008influence,kapferer2009effect,bekki2014galactic,salem2015ram,henderson2016significant,lee2016effect}. \citet{benitez2013dwarf} demonstrated that field dwarfs are also susceptible to these processes when interacting with large-scale filaments of gas. \\
\begin{table*}
    \caption{Properties of simulated dwarf galaxies. Column (2) lists the spline gravitational force softening length for each simulation. Column (3) lists the mass of individual dark matter particles. Column (4) lists the mass of individual star particles at birth. Columns (5) and (6) list the virial and stellar mass, respectively, of each galaxy in our sample at $z=0$. Column (7) lists the number of star particles in each galaxy at $z=0$. Column (8) lists the mass of neutral hydrogen in each of our galaxies at $z=0$. Columns (9) and (10) list the $z=0$ distance from each galaxy to the nearest more massive field galaxy and Milky Way mass galaxy, respectively. Note that while these galaxies are often the host simulation's central halo, they may also be another field galaxy and need not lie in the high resolution region of the simulation. Column (11) lists the type of star formation history that each galaxy possesses. ``Continuous'' SFHs correspond to dwarfs that have experienced star formation throughout the course of the simulation, ``stripped'' SFHs to those that have ceased star formation as a consequence of ram pressure stripping, ``dead'' SFHs to those that stopped forming stars following reionization, and ``gappy'' SFHs to those that undergo intermittent star formation, ceasing to form stars for at least 2 Gyr before resuming star formation.\newline $^\dagger$Backsplash galaxy in the process of falling into the simulation's central galaxy} 
    \label{table:prop}
    \centering
    {\renewcommand{\arraystretch}{1.4}
    \begin{tabular*}{\textwidth}{c @{\extracolsep{\fill}} c c c c c c c c c c}
          \hline
          \hline
          \makecell[c]{Galaxy} 
          & \makecell[c]{$\epsilon$ \\ pc}
          & \makecell[c]{m$_{DM,part}$ \\ 10$^4$ M$_\odot$} 
          & \makecell[c]{m$_{star,part}$ \\ 10$^3$ M$_\odot$} 
          & \makecell[c]{M$_{vir}$ \\ 10$^9$ M$_\odot$}
          & \makecell[c]{M$_{star}$\\ 10$^6$ M$_\odot$}
          & \makecell[c]{N$_{star}$}
          & \makecell[c]{M$_{HI}$ \\ 10$^6$ M$_\odot$}
          & \makecell[c]{D$_{NF}$ \\ kpc}
          & \makecell[c]{D$_{MW}$ \\ Mpc}
          & \makecell[c]{SFH type}
          \\ (1) & (2) & (3) & (4) & (5) & (6) & (7) & (8) & (9) & (10) & (11)\\
         \hline
        \makecell[c]{h239a \\ h239b \\ h239c}
         & \makecell[c]{173}
         & \makecell[c]{13} 
         & \makecell[c]{8.0}
         & \makecell[c]{3.8 \\ 1.0 \\ 0.92}
         & \makecell[c]{0.20 \\ 9.9 \\ 0.069}
         & \makecell[c]{33 \\ 1732 \\ 12}
         & \makecell[c]{14 \\ 0.038 \\ 0.027}
         & \makecell[c]{317 \\ 291$^\dagger$ \\ 269$^\dagger$}
         & \makecell[c]{0.3 \\ 0.3 \\ 0.3}
         & \makecell[c]{gappy \\ stripped \\ gappy}
         
         \\[0.45cm]\makecell[c]{h258a \\ h258b \\ h258c \\ h258d \\ h258e} & \makecell[c]{173}
         & \makecell[c]{13}
         & \makecell[c]{8.0}
         & \makecell[c]{3.5 \\ 2.6 \\ 1.7 \\ 1.2 \\ 0.94}
         & \makecell[c]{57 \\ 1.3 \\ 1.5 \\ 1.3 \\ 3.6}
         & \makecell[c]{9925 \\ 227 \\ 259 \\ 237 \\ 638}
         & \makecell[c]{4.0 \\ 0.11 \\ 1.6$\times$10$^{-5}$ \\ 0 \\ 0.053} 
         & \makecell[c]{482$^\dagger$ \\ 471 \\ 466 \\ 354$^\dagger$ \\ 165$^\dagger$}
         & \makecell[c]{0.5 \\ 0.5 \\ 0.6 \\ 0.4 \\ 0.3}
         & \makecell[c]{continuous \\ dead \\ dead \\ dead \\ dead}
         
         \\[0.85cm]\makecell[c]{h277a \\ h277b \\ h277c \\ h277d} 
         & \makecell[c]{173}
         & \makecell[c]{13}
         & \makecell[c]{8.0} 
         & \makecell[c]{5.5 \\ 2.1 \\ 2.1 \\ 1.2}
         & \makecell[c]{13 \\ 2.9 \\ 0.068 \\ 2.1}
         & \makecell[c]{2293 \\ 512 \\ 12 \\ 366}
         & \makecell[c]{5.5 \\ 0.024 \\ 0 \\ 0}
         & \makecell[c]{407 \\ 286$^\dagger$ \\ 380 \\ 319$^\dagger$}
         & \makecell[c]{2.2 \\ 0.3 \\ 0.4 \\ 0.5}
         & \makecell[c]{continuous \\ stripped \\ dead \\ stripped}
         
         \\[0.61cm]\makecell[c]{h285a}
         & \makecell[c]{173}
         & \makecell[c]{13} 
         & \makecell[c]{8.0} 
         & \makecell[c]{1.7}
         & \makecell[c]{0.064}
         & \makecell[c]{11}
         & \makecell[c]{0.50}
         & \makecell[c]{82}
         & \makecell[c]{0.3}
         & \makecell[c]{gappy}
         
        \\[0.11cm]\makecell[c]{h603a}
         & \makecell[c]{173}
         & \makecell[c]{1.6}
         & \makecell[c]{8.0}
         & \makecell[c]{1.6}
         & \makecell[c]{19}
         & \makecell[c]{3245}
         & \makecell[c]{3.4}
         & \makecell[c]{230$^\dagger$}
         & \makecell[c]{4.5}
         & \makecell[c]{continuous}
         
         \\[0.14cm]\makecell[c]{h986a \\ h986b \\ h986c \\ h986d \\ h986e \\ h986f \\ h986g}
         & \makecell[c]{173}
         & \makecell[c]{13}
         & \makecell[c]{8.0}
         & \makecell[c]{5.8 \\ 4.4 \\ 3.2 \\ 2.1 \\ 1.7 \\ 1.5 \\ 0.98}
         & \makecell[c]{1.5 \\ 6.4 \\ 2.2 \\ 0.47 \\ 0.12 \\ 0.068 \\ 0.068}
         & \makecell[c]{234 \\ 1097 \\ 379 \\ 83 \\ 21 \\ 12 \\ 12}
         & \makecell[c]{17 \\ 7.9 \\ 3.6 \\ 0.015 \\ 0.0043 \\ 4.4$\times$10$^{-5}$ \\ 0}
         & \makecell[c]{396 \\ 272 \\ 365 \\ 249 \\ 62$^\dagger$ \\ 148 \\ 142}
         & \makecell[c]{2.5 \\ 2.1 \\ 2.0 \\ 1.6 \\ 2.0 \\ 2.4 \\ 2.5}
         & \makecell[c]{gappy \\ gappy \\ gappy \\ dead \\ dead \\ dead \\ dead}
         
         \\[1.12cm]\makecell[c]{h516a}
         & \makecell[c]{87}
         & \makecell[c]{1.6}
         & \makecell[c]{0.99}
         & \makecell[c]{3.2}
         & \makecell[c]{0.070}
         & \makecell[c]{98}
         & \makecell[c]{0.024}
         & \makecell[c]{197}
         & \makecell[c]{4.9}
         & \makecell[c]{dead}
        
         \\[0.07cm]\makecell[c]{h2003}
         & \makecell[c]{65}
         & \makecell[c]{0.67}
         & \makecell[c]{0.40}
         & \makecell[c]{8.4}
         & \makecell[c]{20}
         & \makecell[c]{66,266}
         & \makecell[c]{11}
         & \makecell[c]{N/A}
         & \makecell[c]{4.6}
         & \makecell[c]{continuous}
    \end{tabular*}}
\end{table*}
\indent In this work, we focus on a particular type of SFH, characterized by a long period of little to no star formation. The Local Group contains a number of dwarfs that fit this description, including Leo T, Leo A, WLM, and the Aquarius Dwarf Irregular Galaxy \citep{cole2007leo,clementini2012variability,weisz2014star,cole2014delayed}. \citet{benitez2015mergers} investigated a related category -- dwarfs that experience a pronounced dip in star formation, but have approximately equal star formation preceding and following the dip. In their simulations, they found that this was a product of equal-mass mergers bringing in new material. \\
\indent Here, we examine simulated galaxies with gaps of billions of years in their SFHs. Most are initially shut off by cosmic reionization, though a couple experience shut off twice. Both \citet{benitez2015imprint} and \citet{fitts2017fire} found some simulated galaxies shut off by reionization that resumed star formation only after gaining enough halo mass, while \citet{ledinauskas2018reignited} observed a comparable phenomenon in semi-analytic models. Similar scenarios have been proposed to explain the SFHs of a number of Local Group dwarfs. \citet{ricotti2009late} suggests that Leo T experienced a burst of star formation after becoming sufficiently concentrated, while both Leo A \citep{cole2007leo} and Aquarius \citep{cole2014delayed} are thought to have experienced delayed bursts as a result of inefficient gas cooling.\\
\indent In our galaxies, we observe different trends than any of those previously proposed. We find (1) none of our dwarf galaxies undergo equal-mass mergers after $z=3$. (2) Although some experience largely dark mergers with much smaller objects, these do not trigger new star formation. (3) Interactions with the cosmic web or nearby gas ejecta from more massive galaxies compress the gas within our dwarfs to restart star formation. \\
\indent This paper is organized as follows: in Section \ref{sims}, we provide a brief description of our simulations. We present our main results and the cause of reignition of star formation in Section \ref{gals}. In Section \ref{obs}, we explore observational tests of our model with predictions for JWST. We conclude with a summary of our findings in Section \ref{summ}.
\section{The Simulations}
\label{sims}
The dwarf galaxies analyzed in this paper are taken from 8 simulations run using the parallel N-body + smoothed particle hydrodynamic (SPH) code {\sc Gasoline} \citep{Wadsley2004}. Each simulation is a zoomed-in cosmological run within a 25$^3$ or 50$^3$ Mpc$^3$ co-moving box centered on a 10$^{10}$-10$^{12}$ M$_{\odot}$ halo. These central galaxies are selected from a uniform dark matter only simulation, then re-simulated using the zoom-in technique \citep{katz1993hierarchical}. The box is centered upon the chosen galaxy and the particles that end up within 1 Mpc of the main halo at $z=0$ are re-sampled at our highest resolution with the added inclusion of baryons.  The dwarf galaxies described in this work are taken from this high resolution region. \\
\indent Of the 23 dwarfs in our sample, 21 (h239a-c, h258a-e, h277a-d, h285a, h603a, and h986a-g) are selected from 50$^3$ Mpc$^3$ boxes, which have spline gravitational force softening lengths ($\epsilon$) of 173 pc in their high resolution regions. Our remaining two dwarfs are selected from 25$^3$ Mpc$^3$ boxes and are simulated at slightly higher resolutions. h516a has $\epsilon$ $\sim$ 87 pc, while h2003, which is the central galaxy of its simulation, has $\epsilon$ $\sim$ 65 pc. Particle masses, virial masses, stellar masses, neutral hydrogen masses, proximity to other field galaxies, and star formation history types for individual galaxies may be found in Table \ref{table:prop}.\\
\indent Beginning with initial conditions generated from WMAP 3-year data \citep{spergel2007three}, the simulations are permitted to evolve from $z\sim150$ to $z=0$. So as to approximate the effects of reionization, a redshift-dependent UV background turns on at $z=9$, following a modified version of  \citet{haardt1996radiative}, specified in Cloudy \citep{ferland1998cloudy} as ``table HM05''. \\
\indent In order to model gas dynamics, {\sc Gasoline} includes prescriptions for collisional ionization rates \citep{Abel1997}, radiative recombination \citep{Black1981,Verner1996}, photoionization, bremsstrahlung, and cooling from helium, hydrogen - both atomic \citep{Cen1992} and molecular \citep{christensen2012implementing}, and metal lines \citep{Shen2010}. As described in \citet{Stinson2006} and expanded upon by \citet{christensen2012implementing}, star formation is a stochastic process based on the abundance of H$_2$ and the free-fall time of gas that is sufficiently cool and dense. However, while {\sc Gasoline} employs a temperature threshold (T $<$ 1000 K) for star formation, the direct dependence of the star formation recipe upon the H$_2$ fraction makes a density threshold superfluous, ensuring that all stars form at $\rho \gtrsim$ 100 amu cm$^{-3}$. Each cold, dense gas particle has a probability $p$ of forming a star particle:
\begin{equation}
p = \frac{m_{gas}}{m_{star}}(1-e^{c^*\Delta t/t_{form}})
\end{equation}
where $m_{gas}$ is the mass of the original gas particle, $m_{star}$ is the initial mass of the star particle, $\Delta t$ is the star formation timescale (1 Myr in these simulations), $t_{form}$ is the dynamical time, and $c^*$ is a star formation efficiency factor given by 
\begin{equation}
    c^* = 0.1f_{H_2}
\end{equation}
\citep{christensen2012implementing}. Any resultant star particles represent an entire stellar population, with masses in accordance with the initial mass function of \citet{Kroupa1993}. These masses, in combination with metallicities, determine stellar lifetimes as described in \citet{Raiteri1996}. \\ 
\indent Stars with masses between 8 M$_{\odot}$ and 40 M$_{\odot}$ explode as Type II supernovae (SNe), while those that are more massive are assumed to collapse directly to black holes (although it should be noted that we do not model black holes in these simulations). At each timestep, the number of Type II SNe that ought to have gone off in a given star particle is calculated and this number multiplied by 10$^{51}$ ergs injected into the surrounding gas particles. These particles are restricted to be within a ``blastwave'' radius of the supernova \citep{Stinson2006}, which is calculated according to the Sedov-Taylor solution for a blastwave supernova remnant \citep{McKee1977}. Mass and metals are redistributed to the interstellar medium (ISM) with metal diffusion following \citet{Shen2010}. So as to prevent the energy resulting from the SNe from being radiated away too quickly (a potential consequence of the high density regions in which Type II SNe tend to occur),  radiative cooling is also temporarily disabled in the gas particles nearest to the supernova, mimicking the adiabatic expansion phase of the explosion. This not only imitates the impact of the high pressure produced by the expanding supernova by pushing the surrounding gas outward, but also reproduces the effects of the turbulence that results from SNe, briefly limiting further star formation \citep{Stinson2006}.  \\
\begin{figure}
\centering
\includegraphics[scale=0.39]{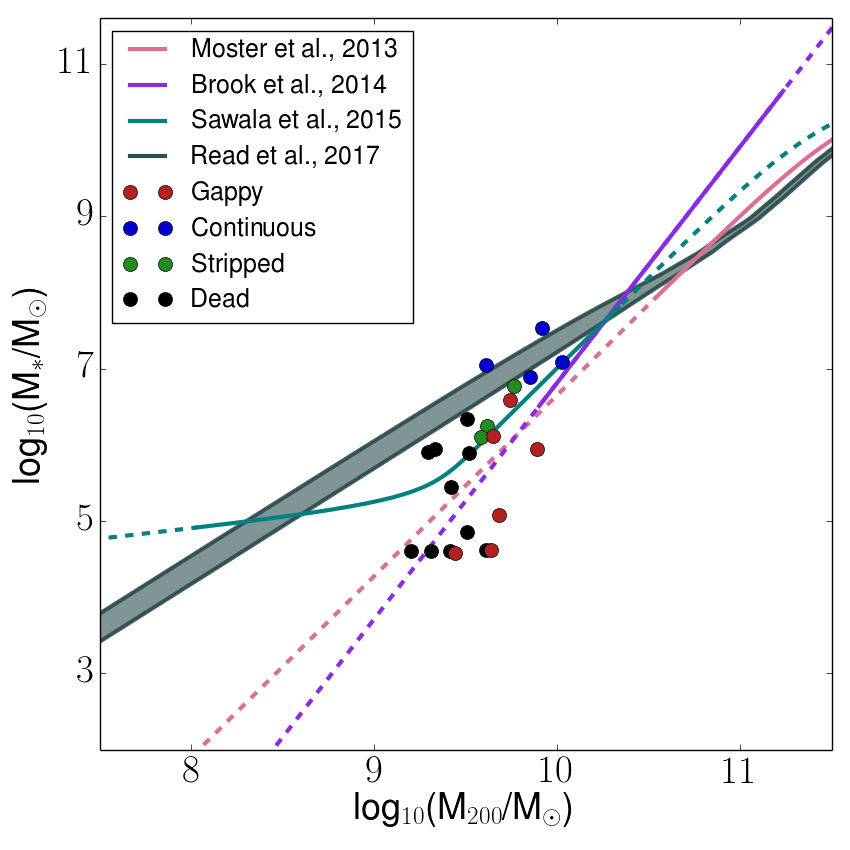}
\caption{Stellar mass - halo mass (SMHM) relation for dwarf galaxies within our sample compared to data from \citet{sawala2015bent} and \citet{read2017stellar} and abundance matching results from \citet{moster2012galactic} and \citet{brook2014stellar}. M$_{200}$ is defined as the total mass contained within the radius at which the mean enclosed density of particles bound to the halo drops below 200$\rho_{crit}$. Dashed lines denote extrapolation. For galaxies that have had their halo masses reduced through stripping, we plot here the maximum halo mass. Stripped galaxies lead to increased scatter in the SMHM relation, but a tighter relation results from using the maximum halo mass \citep{munshi2017going}. Stellar and halo masses are shown as per \citet{munshi2013reproducing}. Although all of the galaxies within our sample lie within a narrow range of halo masses, they occupy a broad span of stellar masses. This is largely due to the fact that those dwarfs that have ``gappy'' or ``dead'' star formation histories tend to form few stars relative to other galaxies of similar mass.}
\label{fig:smhm}
\end{figure}
\indent Approximately 10-20\% of the supernovae in the simulation are Type Ia, originating from less massive ($<$8 M$_\odot$) stars. The number of Type Ia SNe that go off in each star particle is based on the binary fractions from \citet{greggio1983binary}. The feedback from these supernovae is treated in much the same way as that from Type II SNe, except radiative cooling is not disabled. Low mass stars can also return mass to the interstellar medium via stellar winds, following \citet{kennicutt1994past}. These stellar winds from low mass stars account for 99\% of the mass lost by a given star particle over its lifetime, prolonging star formation within the host galaxy \citep{Stinson2006}.\\
\begin{figure*}
\centering
\includegraphics[scale=0.35]{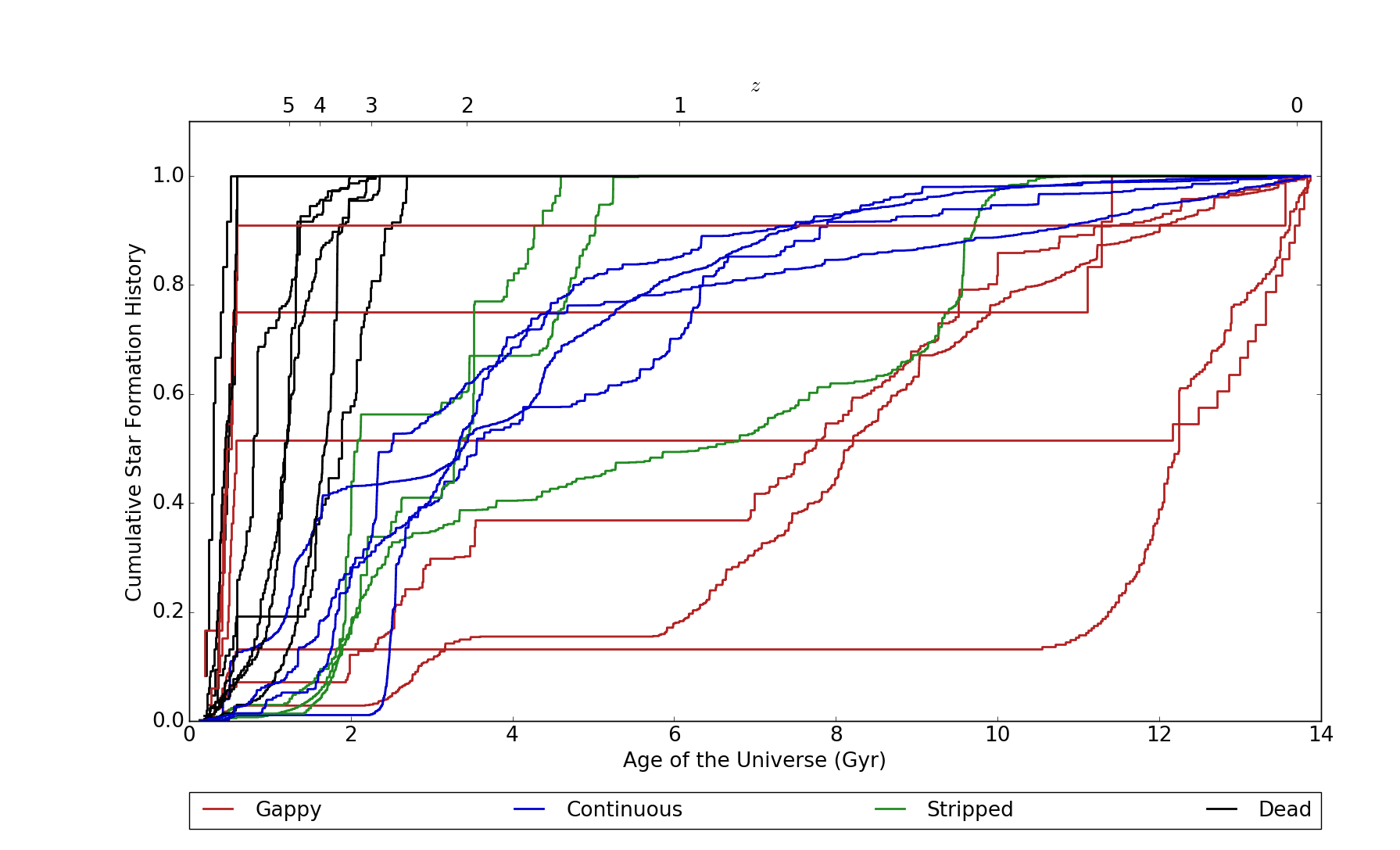}
\caption{Cumulative star formation histories for dwarf galaxies in our sample. From top to bottom at $z=1$, the gappy galaxies are h285a, h239c, h239a, h986c, h986b, and h986a. The lengths of their quiescent periods range from just over 2 Gyr (h986b) to nearly 13 Gyr (h285a).}
\label{fig:CSFH}
\end{figure*}
\indent Haloes are identified and tracked with Amiga's Halo Finder \citep[AHF,][]{knebe2001multi,gill2004evolution}. For each timestep, AHF outputs the halo ownership of each particle in the simulation, allowing us to track a single particle through the entire run and thus compile a detailed accretion and outflow history for any given halo. AHF also calculates the virial mass (M$_{vir}$) of each halo by summing up the masses of the particles contained within the halo's virial radius (R$_{vir}$), including those bound to subhalos. Here, R$_{vir}$ is the radial distance from the center of the halo to the point at which the mean enclosed density of particles bound to the halo drops below $\Delta_{vir}(z)\rho_{b}$. $\Delta_{vir}(z)$ is the virial overdensity, which varies with redshift following \citet{bryan1998statistical}, and $\rho_b$ is the background density. Masses given in this work are virial masses unless otherwise specified. Merger histories for each halo are assembled with {\sc Rockstar} Halo Finder \citep{behroozi2013rockstar} and Consistent Trees \citep{behroozi2013gravitationally}.
\section{Results}
\label{gals}
\begin{figure*}
\centering
\includegraphics[scale=0.35]{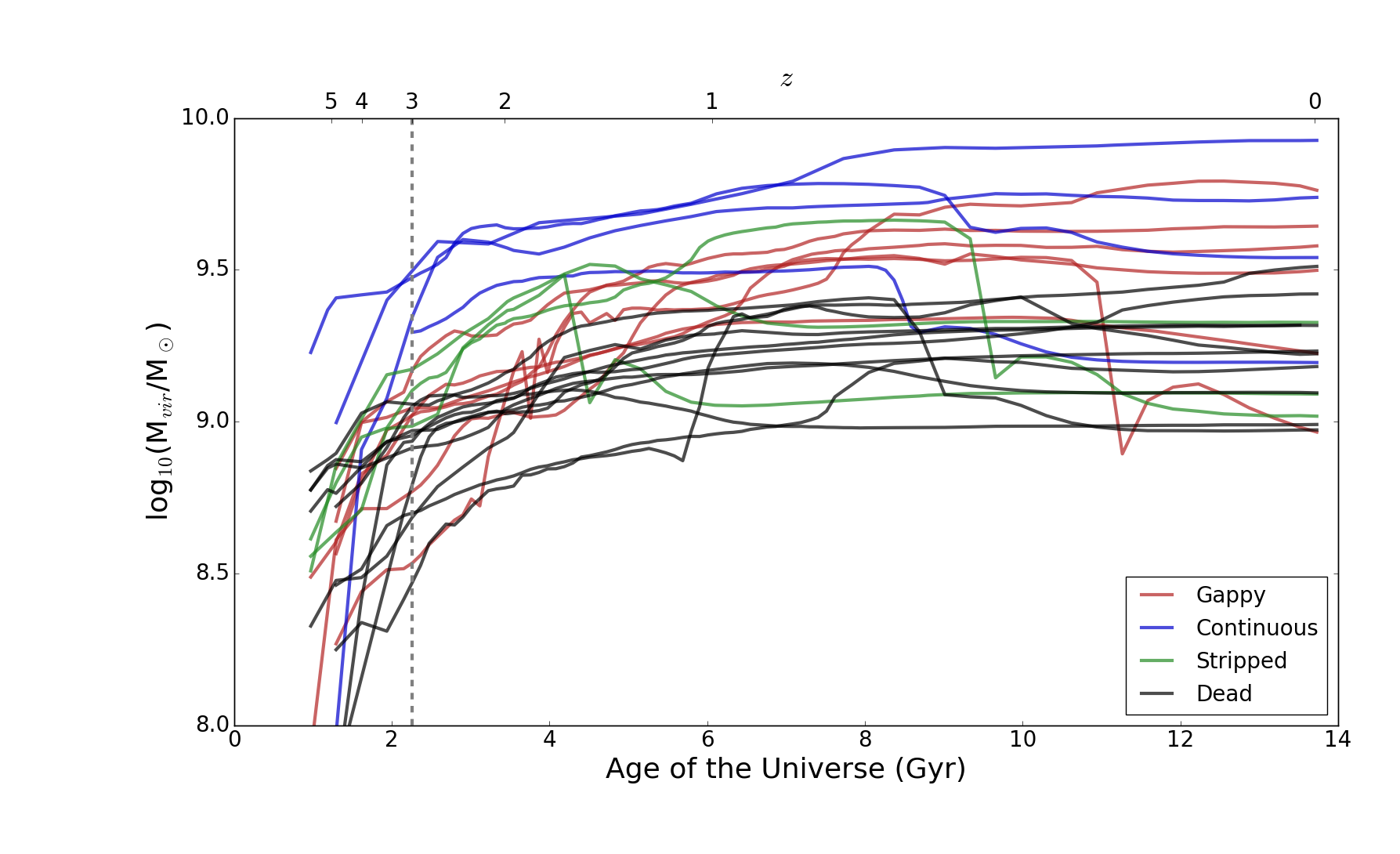}
\caption{Evolution of virial mass over the lifetime of each galaxy in our sample. Those galaxies that are more massive at $z\approx3$ are able to retain their gas through reionization and are thus more likely to continue forming stars. Lower mass halos at $z\approx3$ typically either cease star formation altogether (``dead'' galaxies) or resume it only after an extended gap, placing them in the ``gappy'' category. Note that a number of our galaxies are stripped in halo mass, but are not categorized as ``stripped'' when this interaction does not result in a cessation of star formation. From top to bottom at $z=0$, the gappy galaxies are h986a, h986b, h239a, h986c, h285a, and h239c.}
\label{fig:mvir}
\end{figure*}

Our goal is to study galaxies in which star formation shuts off for an extended period of time, but eventually resumes.  In our simulations, there are 6 galaxies that follow this trend, displaying gaps at $z<3$ of at least 2 Gyr (and as long as 12.8 Gyr) in their star formation histories. All but one are actively forming stars at $z=0$. Henceforth, we will refer to this population as ``gappy'' galaxies. They lie in the halo mass range 9.2$\times$10$^8$ M$_{\odot}$ -- 5.8$\times$10$^9$ M$_{\odot}$, but have formed few stars for their halo mass relative to the other galaxies within our simulation (see Fig \ref{fig:smhm}). In order to examine what properties of galaxies lead to continuous versus non-continuous star formation, we also consider all other field galaxies with 10$^9$ M$_{\odot}$ $<$ M$_{vir}$ $<$ 10$^{10}$ M$_{\odot}$ from the same set of simulations for inclusion in our sample. There are 33 additional galaxies that fit this criteria. We find no halos that are completely dark (zero star particles), but 16 of them have 10 or fewer star particles (due to mass loss resulting from stellar feedback, this corresponds to M$_*$ $\sim$ 6$\times$10$^4$ M$_{\odot}$ for our lowest resolution galaxies), all formed prior to reionization. As it is unlikely that their SFHs are converged, they are excluded from our sample (for a discussion of resolution, see Appendix \ref{appres}). Of the 17 remaining dwarfs, 4 undergo continuous star formation, 10 cease to form stars after cosmic reionization, and 3 experience ram pressure stripping and subsequent cessation of star formation as a result of close encounters with more massive halos post-reionization (see Figure \ref{fig:CSFH} for cumulative star formation histories). \\
\indent The division between galaxies that continue to form stars after reionization and those that do not appears to be based on mass at $z\approx3$ \citep[see also][]{fitts2017fire}, with more massive galaxies able to retain their gas during reionization and thus continue to form stars. This is reflected in Figure \ref{fig:mvir}, which shows the virial mass of each galaxy in our sample as a function of time: galaxies that either cease star formation entirely after reionization, or which do so temporarily, resuming star formation after several Gyr, tend to be the lowest mass galaxies at early times. Those that continuously form stars constitute the highest mass halos at early times. \\
\indent With the aim of determining the cause of the reignition of star formation in the gappy population, we conduct an examination of their accretion histories. In addition to the star formation and virial mass histories shown in Figures \ref{fig:CSFH} and \ref{fig:mvir}, respectively, we track the HI and total gas masses of the galaxies over time, as well as the accretion of gas onto the disks (defined as 0.1R$_{vir}$). These plots are presented in Figure \ref{fig:story}. HI mass most closely tracks star formation within the galaxies in our sample - an unsurprising finding, given that HI formation, like star formation, requires low temperatures and high densities.\\
\indent The total gas masses and gas accretion histories also provide valuable insights into the lives of our dwarfs. For instance, significant decreases in total gas mass which do not coincide with similar depletions of HI indicate that hot gas is being stripped from the halo via some mechanism. Rows 2 and 3 of Figure \ref{fig:story} provide examples of this: as the galaxies draw closer to their simulation's central halo, ram pressure causes some of their gas to be stripped away, leading to a gradual decrease in M$_{gas}$. Over the same span of time, however, M$_{HI}$ remains relatively constant and disk gas accretion, although variable, persists. \\
\indent We also note the timesteps at which our dwarfs merge with halos of at least one-fifteenth their virial mass. At first glance, these mergers may appear to track both HI build-up and resumption of star formation. However, this is really only true in three cases -- h239a and the first reignitions in h986b and h986c (rows 4, 2, and 3, respectively, of Figure \ref{fig:story}). A closer examination of the objects with which h986b, h986c, and h239a merge reveals that they are ultrafaint dwarfs composed almost entirely of dark matter and are thus unable to account for the accretion of gas that succeeds them. For all other gappy galaxies, mergers are either completely absent or occur only at high redshift. We also verified that the gaps were not the product of a merger with a star-rich halo that might have gone undetected in our initial 15-to-1 mass cut. Thus Figure \ref{fig:story} suggests that mergers are not the reason why star formation restarts in these galaxies.
\subsection{Reignition of Star Formation}
Within our sample of dwarf galaxies with resolved SFHs, more than one-fourth have gappy star formation histories. The fact that this is so common suggests either that there are multiple mechanisms capable of reigniting star formation or that there is a single relatively commonplace mechanism at work. In order to gain further insight into what processes might be responsible for restarting star formation, we use the python package {\sc pynbody} to create a series of movies of each halo and its immediate surroundings over the course of the simulation, examining it from different angles and looking at the behavior of each of its components (gas, dark matter, and stars). Stills from three examples are shown in Figures \ref{fig:gasgrid2}, \ref{fig:stream}, and \ref{fig:gasgrid} and are discussed in greater detail in Section \ref{cs1}. \\
\begin{figure*}
\centering
\includegraphics[scale=.76]{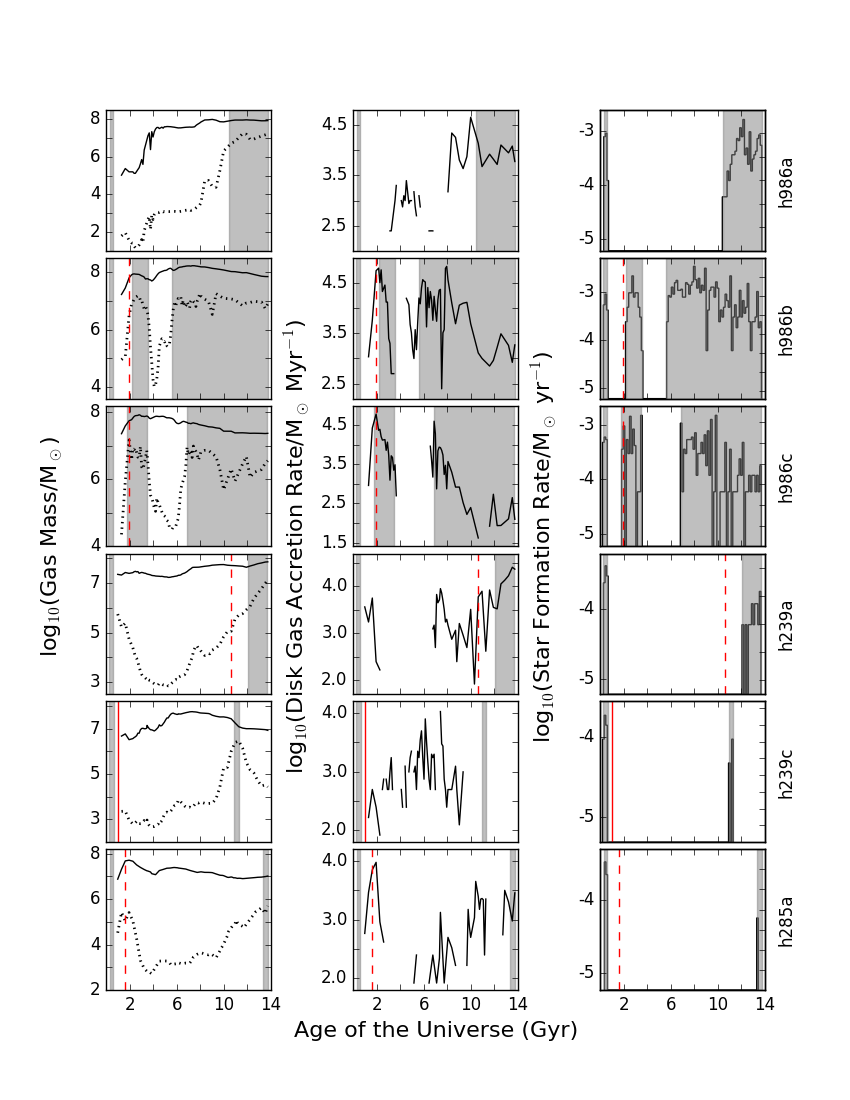}
\caption{\textit{Left}: Total gas mass (solid line) and HI mass (dotted line) of galaxies with gappy star formation histories as a function of time. \textit{Center}: Rate of gas accretion onto disks (defined as 0.1R$_{vir}$) of gappy dwarfs as a function of time. Note that this is calculated only for particles that are entering the disk for the first time. \textit{Right}: Star formation histories of gappy dwarfs. Red lines denote mergers with halos that are at least 1/15th the virial mass of the dwarf at the time of the merger; solid lines indicate major mergers (at least 1:4), while dashed lines indicate minor mergers. Shaded regions correspond to periods of active star formation. For h986a (row 1) and h239c (row 4), total gas mass, disk gas accretion rate, and HI mass consistently correlate with one another. However, discrepancies between these three properties exist for all of the other gappy dwarfs. At low redshift, total gas mass falls while HI mass increases and (in 3 out of 4 cases) disk gas accretion persists, indicating that some mechanism is simultaneously stripping away hot gas and compressing the remaining gas onto the disk to form HI.}
\label{fig:story}
\end{figure*}
\indent We observe that, prior to each star formation reignition event, each dwarf galaxy has a significant interaction with dense gas in the intergalactic medium. In 4 instances, the dwarf passes though a filament of gas independent of the cosmic web of dark matter extending from another halo (usually the central halo). The other 2 dwarfs (both from the same parent simulation) each experience two reignition events. The first is preceded by a high redshift interaction with a filament of gas still associated with dark matter. The second is immediately following a collision with streams of gas thrown off during a nearby merger between two much more massive halos. Although mergers involving the dwarf in question are sometimes seen to accompany these events, they are not the cause of reignition, being neither significant nor consistently present. Rather, they are likely a consequence of our dwarfs entering denser regions of the simulation. \\
\indent Interestingly, these dwarf-dense gas encounters are not necessarily significant accretion events. Although gas is sometimes accreted, this is not the rule (as may be seen in Fig \ref{fig:story}) and, therefore, not the cause of the reignition of star formation. The important factor seems, instead, to be the extent to which the dwarfs' existent gas is compressed as a result of the interaction. This can be determined by calculating the ratio of the ram pressure exerted by the passing stream to the gravitational restoring force of the galaxy itself, where
\begin{equation}
    P_{ram} \propto \rho_{gas}V_{rel}^2
\end{equation}
and
\begin{equation}
    P_{gal} \propto \rho_{gal}V_{200}^2
\end{equation}
\cite[e.g.,][]{benitez2013dwarf}.
Here, V$_{rel}$ refers to the relative velocity between the galaxy and the gas stream with which it interacts, $\rho_{gas}$ is the average density of the portion of the stream through which the galaxy passes, and $\rho_{gal}$ is the average density of the gas contained within R$_{200}$. For each dense gas-dwarf interaction that leads to the reignition of star formation in a galaxy within our simulation, 1 $\lesssim$ $\frac{P_{ram}}{P_{gal}} \lesssim 4$. During periods of quiescence, each dwarf's gas is primarily located in a hot halo surrounding the dwarf. The left column of Figure \ref{fig:story} shows that each of these dwarfs typically contains at least 10$^7$ M$_\odot$ in hot gas within R$_{vir}$. This hot gas, due to its low metallicity and low density, has a cooling time greater than the duration of the simulation \cite[see also][]{christensen2016n}. Under the influence of a moderate amount of ram pressure, however, the gas within the dwarf is compressed onto the disk (with one exception in which it remains in the halo, discussed in Section \ref{cs1}), resulting in densities high enough for HI and H$_2$ to form and thus renewing star formation. \\
\indent It should be noted that encounters with streams of gas are not exclusive to gappy dwarfs. While many of our galaxies are struck by intergalactic streams, those interactions that do not lead to reignition of star formation in dead galaxies are typically characterized by pressure ratios that fall outside of the previously specified range. This is usually due to the characteristics of the stream of gas, rather than those of the galaxy. With the exception of those galaxies that have experienced significant stripping, dead galaxies tend to have P$_{gal}$ values in the same range as those of gappy dwarfs during their quiescent periods.\\
\indent Although ram pressure stripping is the more commonly referenced phenomenon, it is well established that ram pressure is capable of compressing gas, leading to bursts of star formation within galaxies \cite[e.g.,][]{fujita1999effects,bekki2003starbursts,hester2006ram,kronberger2008influence,kapferer2009effect,bekki2014galactic,salem2015ram,henderson2016significant,lee2016effect}. However, in the past, the source of the ram pressure -- the intergalactic medium (IGM) -- has been almost universally modeled as uniform. Even when interactions between dwarfs and the cosmic web have been studied, previous work suggested that such interactions should shut down star formation \cite[e.g.,][]{benitez2013dwarf}, not renew it.  This is the first time that these interactions have been studied in simulations as a trigger for new star formation.
\subsubsection{Case Studies: Gappy Galaxies}
\label{cs1}
\begin{figure*}
\centering
\includegraphics[scale=0.36]{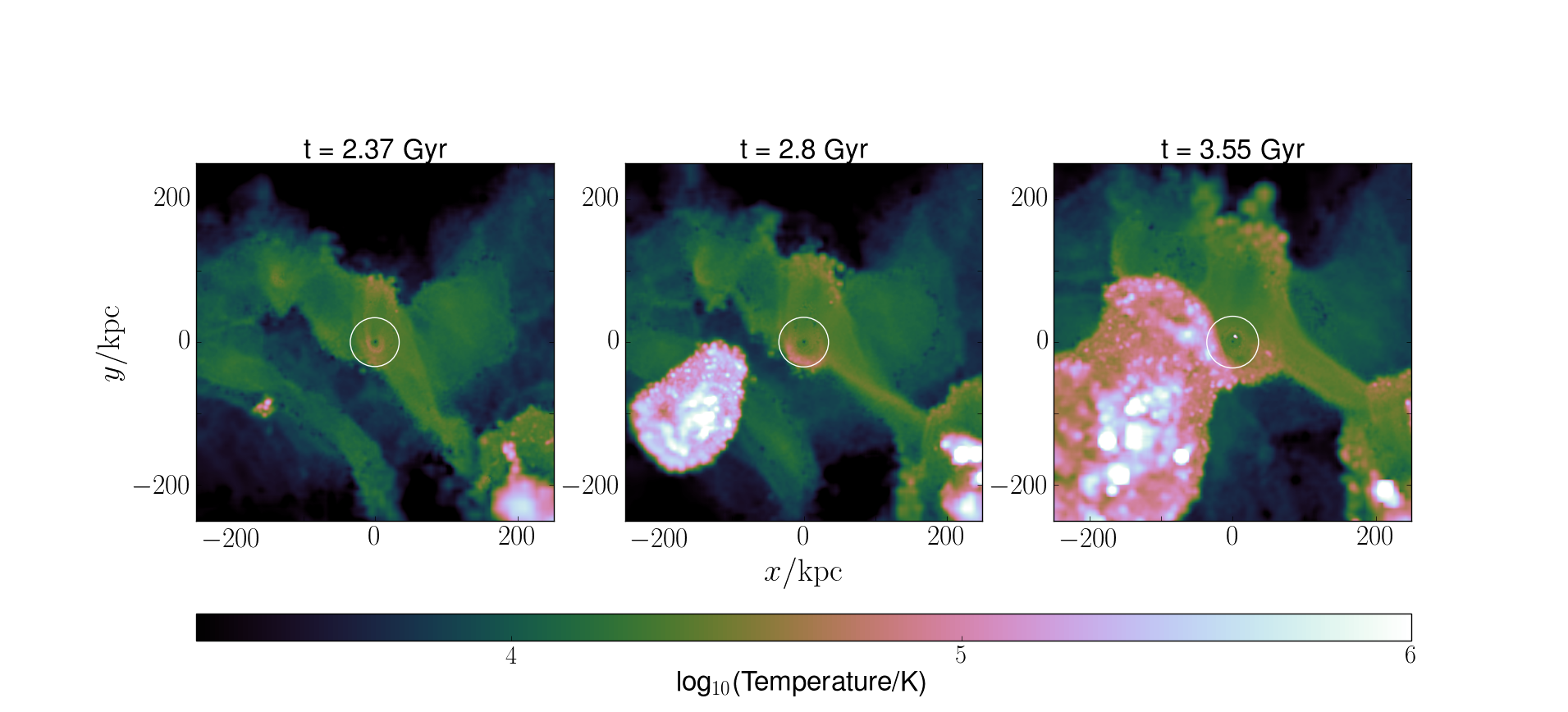}
\caption{Temperature map showing a shock wave from a series of supernovae in a nearby galaxy colliding with h986b. The white circle denotes the virial radius of h986b. The interaction compresses h986b's ISM, allowing it to restore some of the HI reserves depleted through previous star formation and stellar feedback.}
\label{fig:gasgrid2}
\end{figure*}
\begin{figure*}
\centering
\includegraphics[scale=0.6]{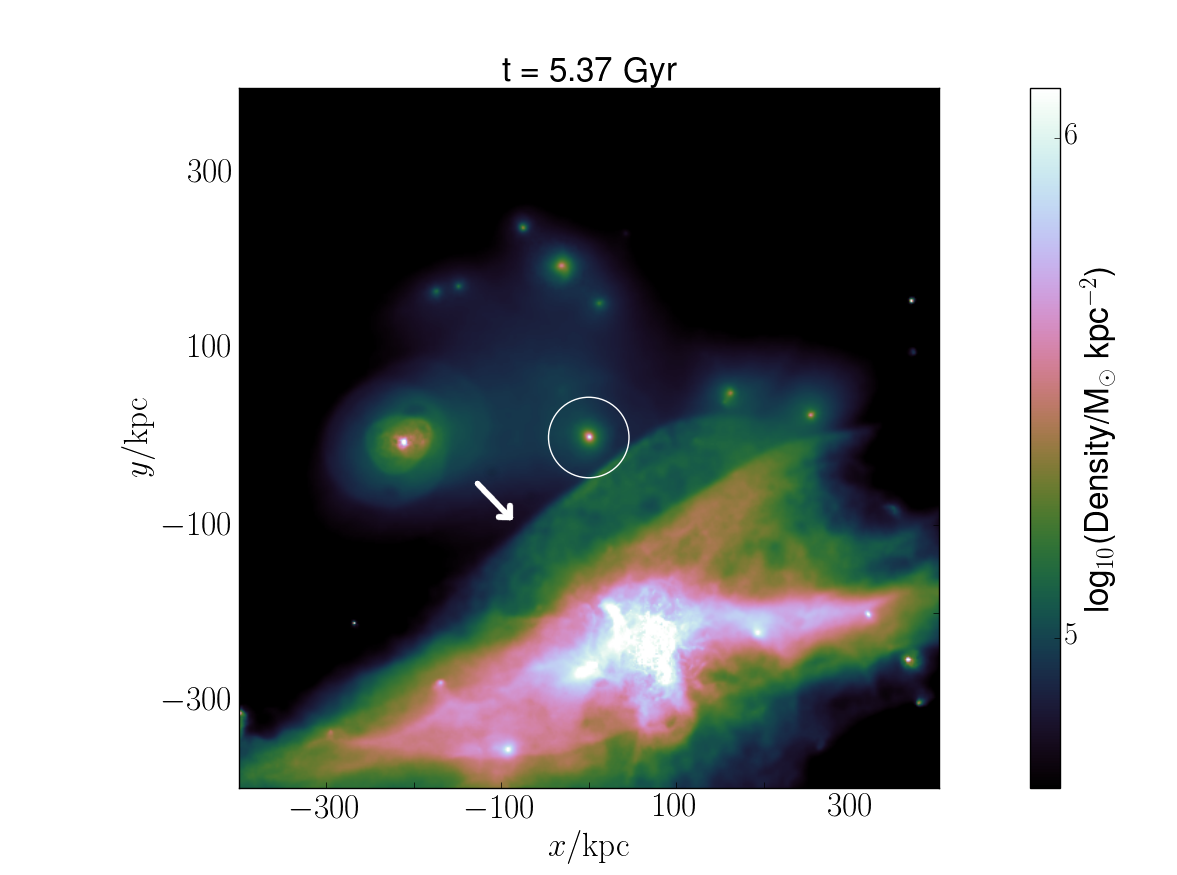}
\caption{Widefield integrated density map showing a stream of gas (indicated by the white arrow) interacting with h986c, which is shown at center. The white circle denotes the virial radius of h986c. The origin of the gas is the merging galaxy in the lower right. The gas compression resulting from the encounter reignites star formation in h986c. A movie of this interaction is available online.}
\label{fig:stream}
\end{figure*}
\begin{figure*}
\centering
\includegraphics[scale=0.36]{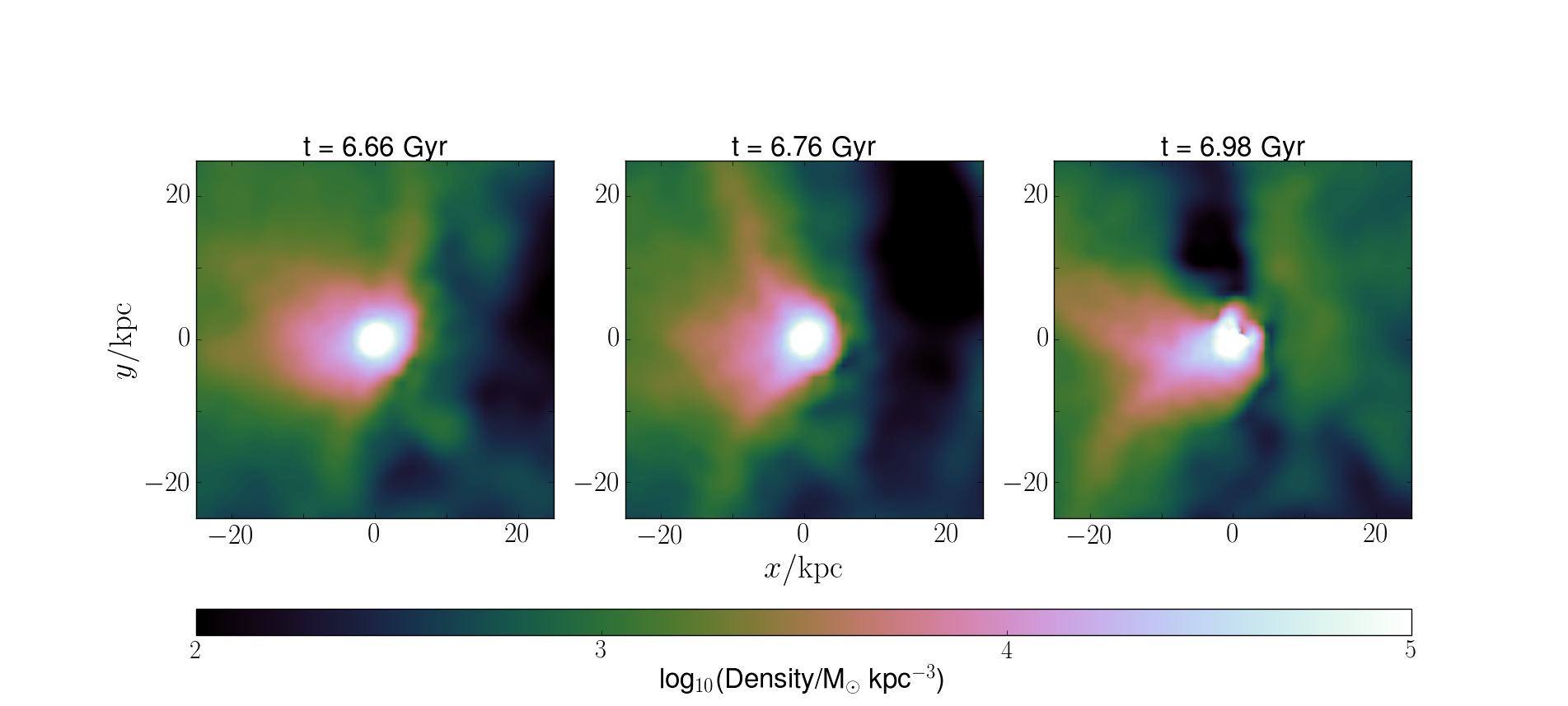}
\caption{Small-scale density slice showing a second stream of gas interacting with h986c. The slightly higher density and velocity of the stream relative to h986c allows it to strip away some of h986c's hot gas, while compressing the galaxy's cooler gas. Consequently, although h986c's total gas mass decreases significantly, its HI mass remains relatively constant and the galaxy is able to continue forming stars. In the third panel, we can see h986c begin to pass through the circumgalactic medium of the simulation's central halo. The ram pressure from this gas will continue the process begun by the earlier stream for the next 5 Gyr, simultaneously stripping away and compressing h986c's ISM. As a result, h986c forms stars for the remainder of the simulation.}
\label{fig:gasgrid}
\end{figure*}

\textbf{h986a}: This galaxy is almost entirely isolated until z$\approx$0.5. Although it begins to rapidly increase in virial mass at t$\sim$8 Gyr, we have verified that there are no mergers at this time.  Rather, the growth appears to be an increase of smoothly accreted mass due to the galaxy entering a more dense environment. At t$\sim$9 Gyr, the dwarf interacts with a stream of gas thrown off by the simulation's central halo (h986, M $\sim 10^{11} M_{\odot}$ at $z=0$). While HI starts to build up at t$\sim$8 Gyr, it rapidly increases again after the stream interaction due to the resulting ram pressure compression (P$_{ram}$/P$_{gal}$ $\approx$ 3). At t$\sim$10 Gyr, h986a acquires a satellite approximately one-sixth of its mass. However, this satellite never merges with the central galaxy, and thus is not indicated as a merger in Figure 4. We verified that no gas particles were transferred from this satellite to the disk of the central galaxy, indicating that the satellite does not contribute to the disk gas accretion evident in row 1 of Figure \ref{fig:story}. The cause of reignition is, therefore, more likely to be the gas compression resulting from h986a's earlier interaction with a stream of gas in the IGM. h986a reaches its minimum distance of 2.1 R$_{vir}$ from h986 at t$\sim$11 Gyr.\\ \\
\textbf{h986b}: This is the first of two galaxies that cease and restart star formation twice over the course of the simulation. Although initially quenched by cosmic reionization, it passes through a filament of gas and dark matter at t$\sim$1 Gyr, accreting a significant amount of gas and building up a reserve of HI (see row 2 of Figure \ref{fig:story}). Star formation resumes at t$\sim$2 Gyr, but is quenched by the galaxy's supernova feedback within 1.5 Gyr. Almost immediately afterward, h986b is struck by a shock wave from a series of supernovae in a neighboring galaxy (see Figure \ref{fig:gasgrid2}). The resulting compression of h986b's gas (P$_{ram}$/P$_{gal}$ $\approx$ 1) leads to an increase in HI mass and the rapid resumption of disk gas accretion that we see in Fig \ref{fig:story}. The process of rebuilding HI reserves is accelerated 1 Gyr later (t$\sim$4.5 Gyr), when h986b is struck by a stream of gas thrown off by h986 during a series of turbulent mergers (P$_{ram}$/P$_{gal}$ $\approx$ 1). Within 1 Gyr, star formation has once again resumed. A second, slightly denser wave of gas succeeds the first, striking h986b at t$\sim$7.5 Gyr (P$_{ram}$/P$_{gal}$ $\approx$ 4), stripping away some of the galaxy's hot gas. h986b reaches its minimum distance of 1.8 R$_{vir}$ from h986 at t$\sim$12 Gyr.\\ \\
\textbf{h986c}: This is the second of two galaxies that cease and restart star formation twice over the course of the simulation. After being quenched by cosmic reionization, it interacts with a filament of gas and dark matter at t$\sim$1.5 Gyr, resuming star formation less than 1 Gyr later. The galaxy remains isolated until t$\sim$5 Gyr, when it is struck by a stream of gas thrown off during a series of mergers between h986 and several smaller neighboring galaxies (see Figure \ref{fig:stream}). Although h986c accretes very little gas during this encounter, the interaction compresses the galaxy's gas (P$_{ram}$/P$_{gal}$ $\approx$ 2) onto its disk, resulting in rapid HI formation (see row 3 of Figure \ref{fig:story}) and a resumption of star formation. A second stream with slightly higher density and velocity and originating from the same source strikes h986c at t$\sim$6.5 Gyr (P$_{ram}$/P$_{gal}$ $\approx$ 3). As can be seen in Figure \ref{fig:gasgrid}, the force from the stream is sufficient to strip away some of the hot gas on the outskirts of the galaxy and compress some of its cooler gas. By t$\sim$7 Gyr, h986c has entered the circumgalactic medium of h986 and will remain within 2 virial radii of the larger halo, where it will undergo mild ram pressure stripping \cite[e.g.,][]{bahe2013does}, for the remainder of the simulation. \\ \\
\textbf{h239a}: This galaxy forms roughly half of its stars at high redshift, but is not massive enough to continue to form stars after reionization. At t$\sim$5 Gyr, h239a is struck by a relatively diffuse stream of gas thrown off during a nearby merger (P$_{ram}$/P$_{gal}$ $\approx$ 0.3), causing it to accrete a significant amount of gas and gradually build up a small reserve of HI. It interacts with a second stream originating from supernovae in a neighboring galaxy 1.5 Gyr later (t$\sim$6.5 Gyr). Although traveling at a similar velocity to the first stream, this stream is slightly more dense (P$_{ram}$/P$_{gal}$ $\approx$ 0.7), resulting in a brief increase in the rate of HI build-up. The fact that neither of these encounters is sufficient to reignite star formation in the galaxy suggests a lower limit of $\approx$ 1 on the P$_{ram}$/P$_{gal}$ value necessary for star formation to restart. At t$\sim$11 Gyr, h239a interacts with a filament of gas (P$_{ram}$/P$_{gal}$ $\approx$ 3) extending off of h239 (M $\sim 10^{12} M_{\odot}$ at $z=0$) as it falls in towards the more massive galaxy. The minor merger that occurs at roughly this same time contributes no gas to h239a, but it may contribute to the compression of the central galaxy in concert with the stream of gas. Within 1 Gyr, star formation has resumed. h239a reaches its minimum distance of 1.5 R$_{vir}$ from h239 at t$\sim$13.5 Gyr.\\ \\
\textbf{h239c}: Almost immediately after being quenched by reionization, h239c experiences a major (2:1) merger. It is relatively isolated until t$\sim$4 Gyr, when it has an encounter with a shock wave from a series of supernovae in a neighboring galaxy (P$_{ram}$/P$_{gal}$ $\approx$ 0.8). The result is significant gas accretion and the formation of HI as gas is compressed onto the disk, as may be seen in row 5 of Figure \ref{fig:story}. Both processes, however, are halted at t$\sim$6 Gyr, when h239c is struck by a filament leftover from a recent merger. The resultant ram pressure (P$_{ram}$/P$_{gal}$ $\approx$ 6) strips away some of h239c's hot gas and even a small amount of its HI. This suggests an upper limit of 4-5 on the P$_{ram}$/P$_{gal}$ value necessary for reignition of star formation. At t$\sim$9 Gyr, h239c passes through a stream of gas blown off by h239. Although the structure is low density, its moderately high velocity (V$_{rel}$ $\approx$ 60 km/s) produces a ram pressure high enough to compress the gas within h239c (P$_{ram}$/P$_{gal}$ $\approx$ 1). The result is the rapid formation of HI that can be seen in Fig \ref{fig:story} and a brief resumption of star formation. In contrast to all of the other gappy dwarfs, however, gas is compressed not onto the disk, but within h239c's halo. This is due to the chaos of the dwarf's immediate surroundings. Over the course of the next 3 Gyr, h239c passes through the simulation's central halo, reaching a minimum distance of 0.2 R$_{vir}$ at t$\sim$11 Gyr. By the time star formation has resumed at t$\sim$11 Gyr, much of h239c's gas has been compressed into a narrow region about 20 kpc from the center of the dwarf. This is why we see significant HI formation, but no disk gas accretion in Fig \ref{fig:story}. Unsurprisingly, this close fly-by eventually rips away most of h239c's gas and much of its dark matter, cutting off star formation.\\ \\
\textbf{h285a}: Between being quenched by cosmic reionization and resuming star formation at t$\sim$13.5 Gyr, this galaxy has 5 encounters with streams of gas in the intergalactic medium. The first two are at t$\sim$3.5 Gyr, when the galaxy is struck by two streams thrown off by neighboring galaxies. Although both interactions have P$_{ram}$/P$_{gal}$ $\approx$ 1, the event leads only to gas accretion and a small amount of HI formation (see row 6 of Figure \ref{fig:story}). Combined with similar occurrences in some of the dead galaxies in our sample, this suggests that (near) simultaneous interactions create too turbulent an environment for star formation to restart. h285a is struck by a third stream from a series of nearby supernovae at t$\sim$7 Gyr (P$_{ram}$/P$_{gal}$ $\approx$ 0.9), leading to a small boost in HI mass. The dwarf encounters successive streams thrown off by h285 (M $\sim 10^{12} M_{\odot}$ at $z=0$) at t$\sim$9 Gyr (P$_{ram}$/P$_{gal}$ $\approx$ 3) and t$\sim$12.5 Gyr (P$_{ram}$/P$_{gal}$ $\approx$ 4) as it falls in towards the larger halo, resulting in a rapid build-up of HI and the formation of a single star particle in the final timestep of the simulation. Because we can't be sure that star formation would continue past $z=0$, we have marked h285a as a marginal case in Figures \ref{fig:mag} and \ref{fig:locvol} (unfilled gappy symbol). For further discussion of this, see Appendix \ref{appres}. h285a reaches its minimum distance of 1.3 R$_{vir}$ from h285 at t$\sim$13.5 Gyr.\\
\subsection{Comparison to Previous Work}
\label{comp}
Other groups have also studied the influence of the cosmic web on star formation in dwarf galaxies. Of particular relevance is \citet{benitez2013dwarf}, which discusses a Local Group simulation in which a population of isolated dwarfs (hereafter referred to as the CLUES dwarfs) pass through a large pancake of gas that develops at $z\sim2$. In contrast to the ram pressure compression that we see in our simulations, this encounter results in ram pressure stripping and the subsequent near-termination of star formation in a number of dwarfs. This is, however, largely unsurprising given the differences between the interactions described in \citet{benitez2013dwarf} and those that we have reported in this paper. Likely of greatest import is the much higher relative velocity that exists between the CLUES dwarfs and the pancake of gas. While our dwarfs are of comparable mass, the CLUES V{$_{rel}$} is up to an order of magnitude higher than that which we typically observe. Accordingly, the resultant P$_{ram}$/P$_{gal}$ value is significantly larger -- well within the range where we would expect pure stripping to occur. Additionally, the interactions that we have described largely concern streams of gas (many from chaotic merger events or star formation in nearby galaxies rather than from the cosmic web), instead of pancakes, and therefore likely involve a smaller total volume of gas, making ram pressure stripping less effective. \\
\indent CLUES has also been used to study gappy dwarfs. \citet{benitez2015imprint} finds that star formation is disfavored at intermediate times (4$<$t/Gyr$<$8) in nearby dwarf galaxies and notes that, in their simulations, `two-component' systems are the product of late merger events. This population is further explored in \citet{benitez2015mergers}, in which the authors demonstrate that the roughly equal-mass mergers occur between galaxies just massive enough to retain some of their gas past reionization. The violence of the interaction disperses the initial stellar populations of the dwarfs while compressing their gas, allowing the product of the merger to form new stars. Consequently, the $z=0$ stellar population is composed of young, relatively metal-rich stars that are centrally concentrated and old metal-poor stars that tend to lie farther from the galactic center. \\
\indent Our gappy dwarfs differ from these two-component systems in a number of ways. Most significantly, as discussed in Section \ref{gals}, our galaxies do not resume star formation as a consequence of gas-rich mergers. There are only three instances in which mergers coincide with star formation reignition (see rows 2, 3, and 4 of Fig \ref{fig:story}), and all are with small objects composed almost entirely of dark matter. Indeed, none of the dwarfs in our sample (including those that have more mundane star formation histories) experience major mergers after $z\sim3$ \citep[see also][]{fitts2018dwarf}. Although we see a small age-metallicity gradient in our gappy dwarfs, it is not nearly as dramatic as that shown in \citet{benitez2015mergers}. This makes sense: it is unlikely that the ram pressure that causes our dwarfs to form new stars would disturb the older stellar populations nearly as significantly as the mergers that \citet{benitez2015mergers} describes. It is possible that this difference in observed stellar population characteristics may provide a way to distinguish between the two scenarios. However, the complexity of galactic evolution complicates this. Our star formation recipe, unlike that used in \citet{benitez2015mergers}, is capable of forming dark matter cores. \citet{el2016breathing} showed that this process can reshuffle stars, pushing old stars outward and thereby creating an age-metallicity gradient. Further work is required to determine whether or not mergers and core formation produce distinguishable gradients.\\
\indent Another difference between our samples lies in the selection criteria. \citet{benitez2015mergers} specifically identified galaxies with little to no intermediate age stars, but with roughly equal star formation at late and early times. By contrast, our sample was chosen purely on the basis of the existence of a significant gap in star formation at any time -- we even have two gappy galaxies with considerable star formation at intermediate times. With the possible exception of h239a, the remainder of our gappy dwarfs are weighted towards either an old population or a young one. These differences may account for our different results. \\
\begin{figure*}
\centering
\includegraphics[scale=0.35]{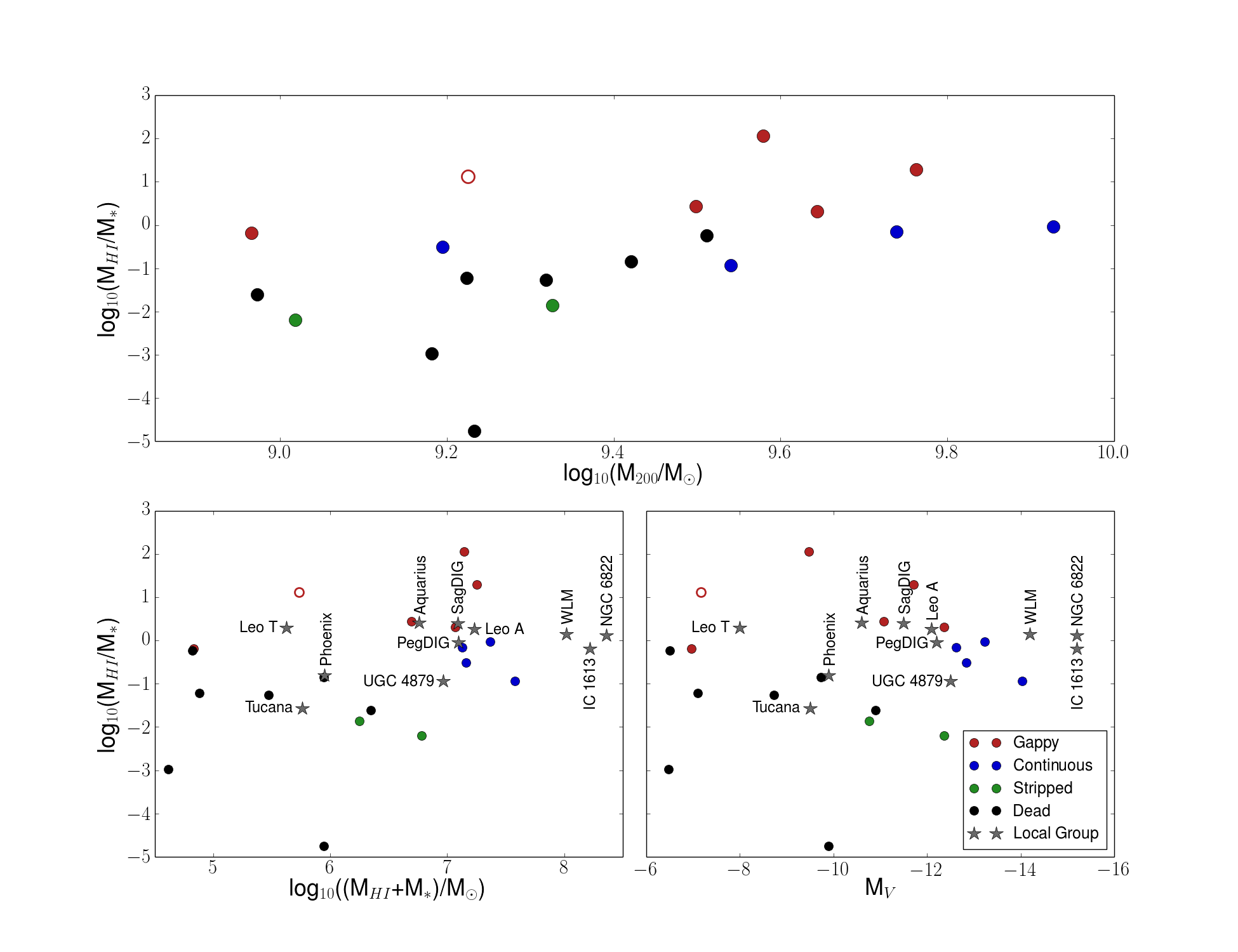}
\caption{Ratio of HI mass to stellar mass plotted against M$_{200}$ (top panel), M$_{HI}$+M$_*$ (bottom left panel), and V-band magnitude (bottom right panel) for dwarf galaxies within our sample and 11 isolated Local Group dwarfs. Values for the latter are from \citet{mcconnachie2012observed}. The unfilled red circle is h285a. Because the gappy dwarfs tend to be galaxies that have formed few stars (and thus have low stellar masses) and have built up a significant amount of HI as a consequence of ram pressure compression, they typically have relatively high ratios of M$_{HI}$ to M$_*$. This provides us with an observational diagnostic with which to separate them out from the other dwarfs in our sample. We can also use this trend to predict which LG dwarfs might have gaps in their star formation histories. As in Fig \ref{fig:smhm}, stellar masses are calculated based on their photometric colors as in \citet{munshi2013reproducing}. Note that simulated dwarfs that contain no HI are not shown.}
\label{fig:mag}
\end{figure*}
\indent A potential relative of our gappy dwarfs appears in \citet{shen2014baryon}, which examines isolated field dwarfs in a cosmological simulation run with a variation on the formulation of {\sc Gasoline} used in this work. Two of their six simulated dwarfs have extremely delayed star formation, experiencing their initial burst only after the simulation has been run for 9-11 Gyr. This delay, however, seems rather unphysical in a group environment -- every dwarf in the Local Group experiences at least some star formation prior to reionization \citep{brown2012primeval,weisz2014star,brown2014quenching}. \citet{shen2014baryon} note that, in both cases, star formation begins following an interaction of some sort: 1 dwarf merges with a dark halo, while the other passes very close to a larger halo and experiences some stripping (similar to our h239c). It seems possible that, with different physics or in different environs, these dwarfs might have gappy SFHs, rather than delayed ones. \citet{shen2014baryon} does use a stronger UV background \citep{haardt2012radiative} than we do, but a re-simulation in which reionization was not included did not alter the delayed SFHs. It remains to future work to determine which conditions lead to early star formation (and thus the possibility of gappy galaxies) and which are conducive to delayed star formation.
\section{Observational Signatures}
\label{obs}
\begin{figure*}
\centering
\includegraphics[scale=0.35]{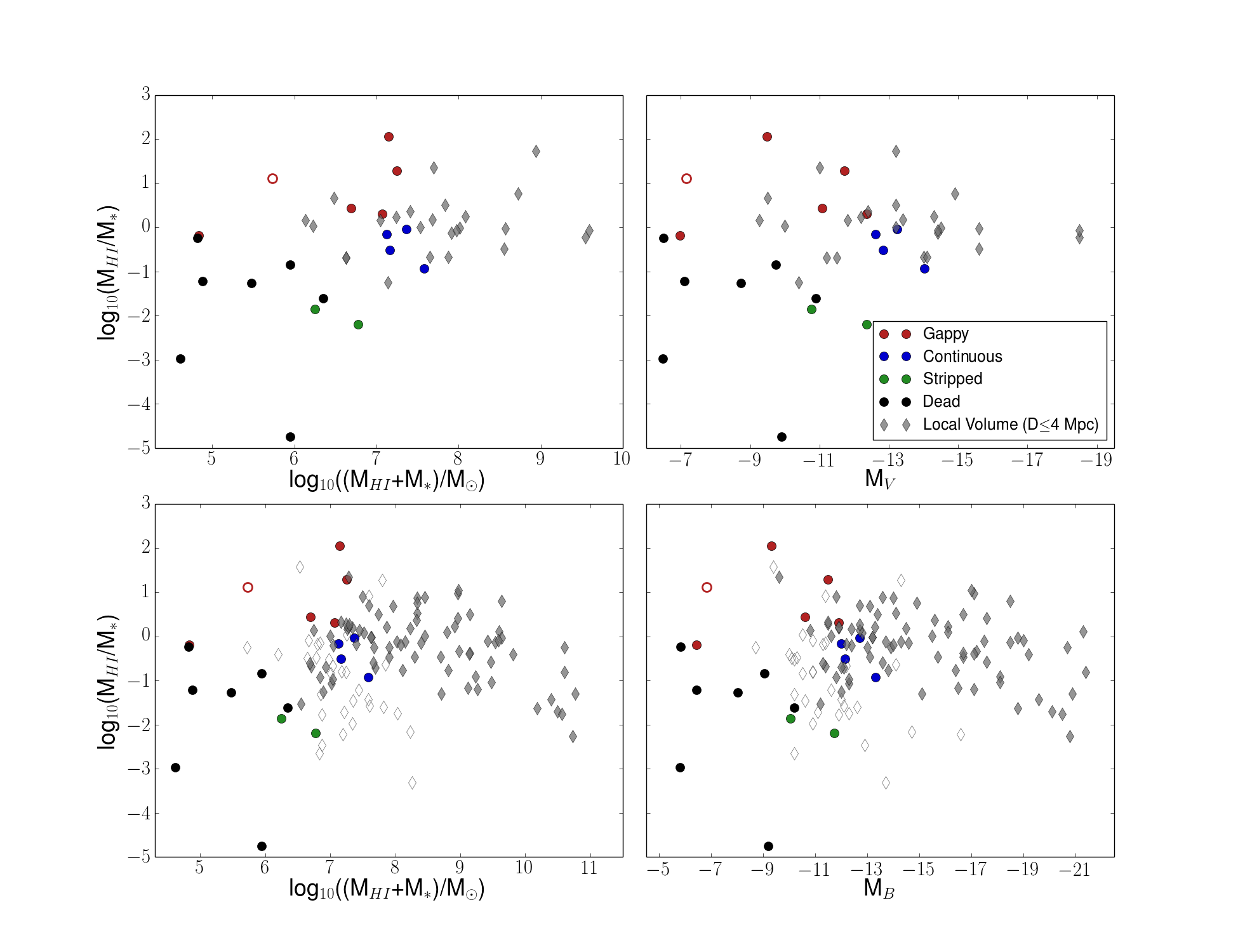}
\caption{Ratio of HI mass to stellar mass plotted against M$_{HI}$+M$_*$ (left panels), V-band magnitude (upper right panel), and B-band magnitude (bottom right panel) for dwarf galaxies within our sample and Local Volume galaxies. Values for the upper two panels are taken from \citet{mcconnachie2012observed} and \citet{mcquinn2015leo} and correspond to galaxies that lie within 3 Mpc of the Milky Way. Values for the bottom two panels are taken from \citet{karachentsev2013updated}, with stellar masses calculated in accordance with \citet{lelli2016sparc}, and include galaxies that are as far as 4 Mpc away. Unfilled diamonds indicate galaxies for which an upper HI mass limit has been used. The unfilled red circle is h285a. As in Fig \ref{fig:mag}, we have excluded galaxies that are known to be satellites.}
\label{fig:locvol}
\end{figure*}
\indent Star formation histories based on resolved color magnitude diagrams (CMDs), like those presented in \citet{weisz2011acs,weisz2014star} and \citet{cole2007leo,cole2014delayed}, are currently the only way to unambiguously identify gappy galaxies in the real universe. However, these are incredibly time-consuming and difficult to obtain. With the aim of finding a simpler method by which potentially gappy galaxies might be identified, we examine the more easily observable features that distinguish the gappy galaxies from the rest of our simulated dwarfs. \\
\indent In most respects, gappy galaxies very much resemble other galaxies of similar stellar mass. There are no offsets in either stellar or gas-phase metallicity between the various groups of simulated galaxies, beyond that which naturally arises from the higher stellar masses of the continuous group. All of the galaxies within our sample follow the observed relations of \citet{kirby2013universal} and \citet{lee2006extending}. However, we find that the galaxies that both cease and resume star formation tend to have higher HI to stellar mass ratios at $z=0$ than those that either form stars continuously throughout the simulation or become quiescent due to reionization or ram pressure stripping. This is primarily due to a combination of two factors: (1) these are galaxies that have built up a significant mass of HI as a consequence of ram pressure compression. (2) These are galaxies that have formed comparatively few stars for their mass -- even at $z=0$, they are not undergoing starbursts, forming only 10$^{-4}$--10$^{-3}$ M$_\odot$yr$^{-1}$ (similar to galaxies in the continuous group) despite their extra HI fuel. Accordingly, we can use M$_{HI}$/M$_*$ to distinguish these galaxies from the rest of our sample. \\
\indent As may be seen in the top panel of Fig \ref{fig:mag}, plotting M$_{HI}$/M$_*$ against halo mass effectively separates halos that have restarted star formation from those that have not. However, although the halo mass of real galaxies can be estimated, it is not, strictly speaking, an observable. Therefore, we also plot M$_{HI}$/M$_*$ against the sum of HI and stellar mass (bottom left panel) and against V-band magnitude (bottom right panel). The gappy dwarfs have systematically higher M$_{HI}$/M$_*$ ratios. \\
\indent There is one dead dwarf galaxy with a M$_{HI}$/M$_*$ ratio in the same range as the gappy dwarfs (see black point at far left in bottom panels and center in top panel of Fig \ref{fig:mag}). This is a single galaxy (h516a) that forms stars only prior to reionization. An examination of its history reveals that the galaxy is struck by a wave of gas from a nearby merger at t$\sim$10.5 Gyr and is building up HI when the simulation terminates. It seems likely, then, that this galaxy would resume star formation in the near future if the simulation were run past $z=0$. h516a is discussed further in Appendix \ref{appres}.\\
\indent By construction, when a galaxy restarts star formation after a lengthy hiatus, its star formation rate (SFR) at that time will be above its past average SFR. This definition may also be used to define a starburst, and a starbursting galaxy may also have a high M$_{HI}$/M$_*$ ratio. We note, however, that even our gappy galaxies that restart star formation at $z\sim1$ continue to maintain a high M$_{HI}$/M$_*$ ratio even at $z=0$, long after their bursts would be complete. Nevertheless, it is not clear that a high M$_{HI}$/M$_*$ ratio must be uniquely indicative of a gap rather than some other process that leads to an increase in M$_{HI}$ and a burst of star formation. Thus, below we discuss in detail the observed galaxies that have high M$_{HI}$/M$_*$ and have had their SFHs studied using resolved stellar populations, in order to verify whether a high M$_{HI}$/M$_*$ ratio does indeed seem to correlate with a gap in star formation. \\
\indent In order to identify nearby dwarfs that might have gaps in their star formation histories, we have plotted 11 isolated Local Group (LG) dwarfs alongside our simulated dwarfs in the bottom panels of Fig \ref{fig:mag}. To ensure that a fair comparison is being made, we have selected only those galaxies that lie outside of the virial radius of the Milky Way and Andromeda, but which are located within 1500 kpc of the center of the Local Group (1500 kpc being roughly the size of the high resolution region around a central galaxy that our simulated dwarfs are drawn from, and,  conveniently, approximately the distance to which resolved stellar population work can be done with the Hubble Space Telescope). The selected dwarfs have stellar masses ranging from  10$^5$ M$_\odot$ to 10$^8$ M$_\odot$, which compares well with our simulated dwarfs.\\ 
\indent Interestingly, the LG dwarfs largely fall along the boundary between the gappy population and the other simulated dwarfs. However, there are a few that stand out. Leo T is quite firmly in the gappy zone and, even given the uncertainties, the star formation history presented in \citet{weisz2014star} suggests the existence of a gap of 3--5 Gyr. The same gap also appears in the SFH presented in \citet{clementini2012variability}, although it is slightly less pronounced, with star formation rates consistent with 0 M$_\odot$ yr$^{-1}$ for $\sim$4 Gyr. Phoenix and Tucana are unambiguously placed in the non-gappy region. The small uncertainties in the \citet{hidalgo2009extended} and \citet{weisz2014star} SFHs of the former leave little room for any sort of cessation of star formation. \citet{monelli2010acs}'s SFH of the latter shows that Tucana is an old galaxy, having formed no stars in the last 8-9 Gyr. This is reflected in both galaxies' positions in Fig \ref{fig:mag}.\\
\indent IC 1613, Leo A, and the Aquarius Dwarf Irregular galaxy are all located considerably closer to the border between the gappy and non-gappy galaxies in Fig \ref{fig:mag}. IC 1613 is shown to have continuously formed stars throughout its lifetime in the SFH presented by \citet{skillman2014acs}. Leo A, however, appears to have a significant star formation hiatus in the SFHs of both \citet{cole2007leo}, which shows a gap of $\sim$3 Gyr, and \citet{weisz2014star}, which shows a gap of 3--6 Gyr. Aquarius is a little more ambiguous. Although \citet{weisz2014star} indicates that Aquarius likely has a gap on the order of 4 Gyr (potentially followed by a second, slightly shorter gap), the somewhat better constrained SFH presented by \citet{cole2014delayed} suggests a period of little to no star formation for no longer than $\sim$2 Gyr - the lower limit of the gaps that we see in our simulated dwarfs. Its relatively high M$_{HI}$/M$_*$ ratio is, then, consistent with our model, but it might be worth further study. All three galaxies have current star formation rates comparable to the gappy galaxies from our sample, so there are no obvious indications that they may be starbursting \citep{mateo1998dwarf, cole2007leo, cole2014delayed}.\\
\indent UGC 4879 presents an interesting case. Despite its low M$_{HI}$/M$_*$ ratio, there is evidence to suggest that it has little to no star formation at intermediate times, but significant populations of both young and old stars \citep{jacobs2011star}. However, it should be noted that the CMDs used to construct the SFHs that support this conclusion do not reach the oldest main sequence turn-off. A deeper CMD is really needed to confirm these findings. \\
\indent Of the remaining LG dwarfs, SagDIG, WLM, and PegDIG seem like the most likely gappy candidates based on their Weisz et al. SFHs, their placement on our plots, and the fact that their current SFRs are within the range of our gappy sample \citep{gallagher1998wide, dolphin2000hubble, hunter2004star}. However, the uncertainties in their SFHs are too significant for a definitive statement to be made. Accordingly, they might be good candidates for further observations.\\
\indent Although the more distant galaxies of the Local Volume are largely outside the range at which a resolved CMD can be acquired with the Hubble Space Telescope, they ought to be within the scope of the James Webb Space Telescope (JWST). In anticipation of the launch of JWST, we have reproduced the lower panels of Fig \ref{fig:mag} with galaxies out to 4 Mpc in Fig \ref{fig:locvol}, using data from \citet{mcconnachie2012observed}, \citet{mcquinn2015leo}, and \citet{karachentsev2013updated}. In the interest of observing overall trends, we have not made any mass cuts, plotting every field galaxy within 4 Mpc for which an HI measurement has been made. Because we lack galaxies within our sample that are at the higher end of this mass range, we will not make predictions about why certain massive galaxies within the Local Volume might be gas rich (although they could be interesting subjects for further study). However, there are a number of dwarf galaxies that lie in the gappy zones of Fig \ref{fig:locvol}. Of particular interest are DDO 113 and KKR 3, which are located in the upper two panels, and BK3N and UGCA 292, which are located in the lower two panels. All 4 of these galaxies have M$_V$ $\geq$ -11 (or M$_B$ $\geq$ -12) and $\frac{M_{HI}}{M_*}$ $\geq$ 4.5. Although each looks as if it might be gappy in the SFHs presented in \citet{weisz2011acs}, the uncertainties are simply too large for any conclusions to be drawn. However, given their relative proximity to the Milky Way, they would be ideal candidates for JWST observations.
\section{Summary}
\label{summ}
In this work, we have used high resolution cosmological simulations to study the star formation histories of dwarf galaxies in group environments. In particular, we have investigated a population of galaxies in which star formation has ceased for at least 2 Gyr before being reignited via some mechanism. Our primary findings are as follows:\\
\indent (i) Nearly 20\% of the isolated dwarf galaxies in our simulations with masses between 10$^9$ M$_\odot$ and 10$^{10}$ M$_\odot$ cease and resume star formation at least once over the course of the simulation. This suggests that the mechanism responsible for the reignition of star formation must be relatively commonplace. \\
\indent (ii) Our dwarfs resume star formation as a result of interactions with streams or filaments of gas originating from nearby mergers or the cosmic web. Where the ratio of the ram pressure exerted by the external gas to the gravitational restoring force of the galaxy itself is between approximately 1 and 4, the gas within the galaxy is compressed, leading to the formation of HI and, eventually, new stars.\\
\indent (iii) Because dwarf galaxies within this population tend to have built up a large store of HI but formed few stars, we can use the ratio of their HI masses to their stellar masses to distinguish them from other isolated dwarfs in the same mass range. By comparing the resultant values to those of similar dwarfs in the nearby universe, we can predict which Local Volume dwarfs might have gaps in their star formation histories. \\
\indent Isolated field dwarfs are rarely truly isolated throughout their histories \citep{2017arXiv171110620L}. Mergers -- both major and minor, encounters with other halos, and interactions with the filaments of gas and dark matter that span the intergalactic medium all leave their imprints on dwarf galaxies. We have identified a possible signature of some of this chaotic history in the star formation histories of our simulated dwarfs, as well as those of several Local Group and Local Volume dwarfs.
\section*{Acknowledgments}
 ACW is supported by a Government Assistance in Areas of National Need grant, the Noemie Koller Endowed Scholarship, and an ACM SIGHPC/Intel Computational \& Data Science fellowship. ACW and AMB are supported by Hubble Grant HST-AR-14281. DRW is supported by a fellowship from the Alfred P. Sloan Foundation. Resources supporting this work were provided by the NASA High-End Computing (HEC) Program through the NASA Advanced Supercomputing (NAS) Division at Ames Research Center. We thank Fabio Governato, Tom Quinn, Sijing Shen, and James Wadsley for use of the {\sc Gasoline} code and help with initial conditions. We also thank the anonymous referee for their comments, which helped to improve the content and form of this paper. The python packages {\sc matplotlib} \citep{Hunter2007}, {\sc numpy} \citep{walt2011numpy}, and {\sc pynbody} \citep{pynbody} were all used in parts of this analysis.
\bibliographystyle{mnras}
\bibliography{DwarfSFReignition_AWright_v2}
\appendix
\section{Resolution}
\label{appres}
\begin{figure}
\centering
\includegraphics[scale=0.37]{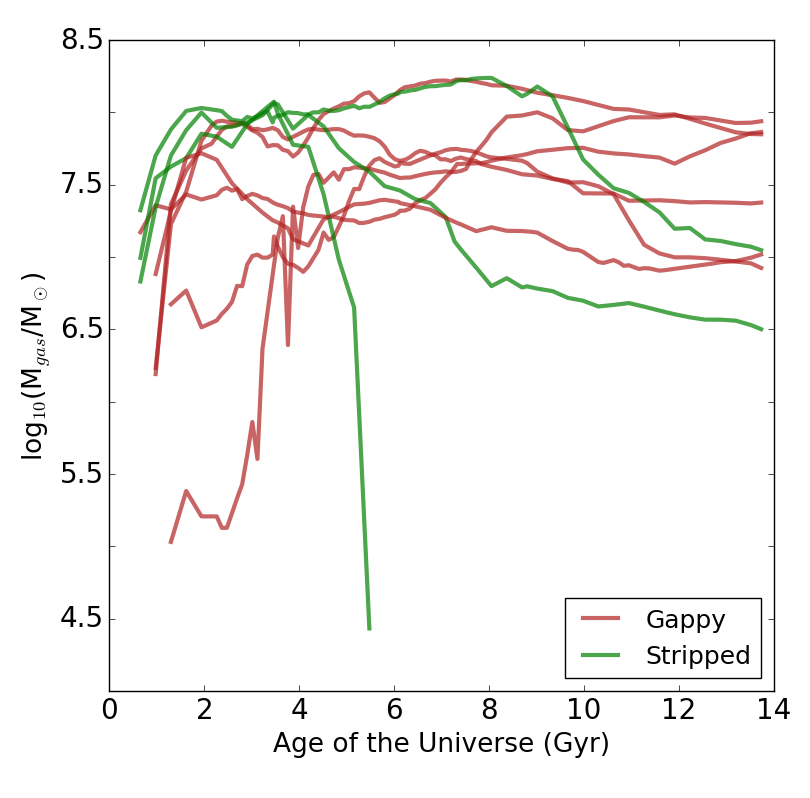}
\caption{The gas mass of each stripped or gappy galaxy in our sample as a function of time. Prior to the stripping of the former group, both types of galaxies share similar M$_{vir}$ (see Figure \ref{fig:mvir}) and M$_{gas}$, even though the stripped galaxies form hundreds to thousands of star particles, while the gappy galaxies form as few as 11. Therefore, the gappy galaxies are not quiescent for lack of gas -- there is some other process that is preventing them from forming stars.}
\label{fig:mgas}
\end{figure}
\begin{figure*}
\centering
\includegraphics[scale=0.39]{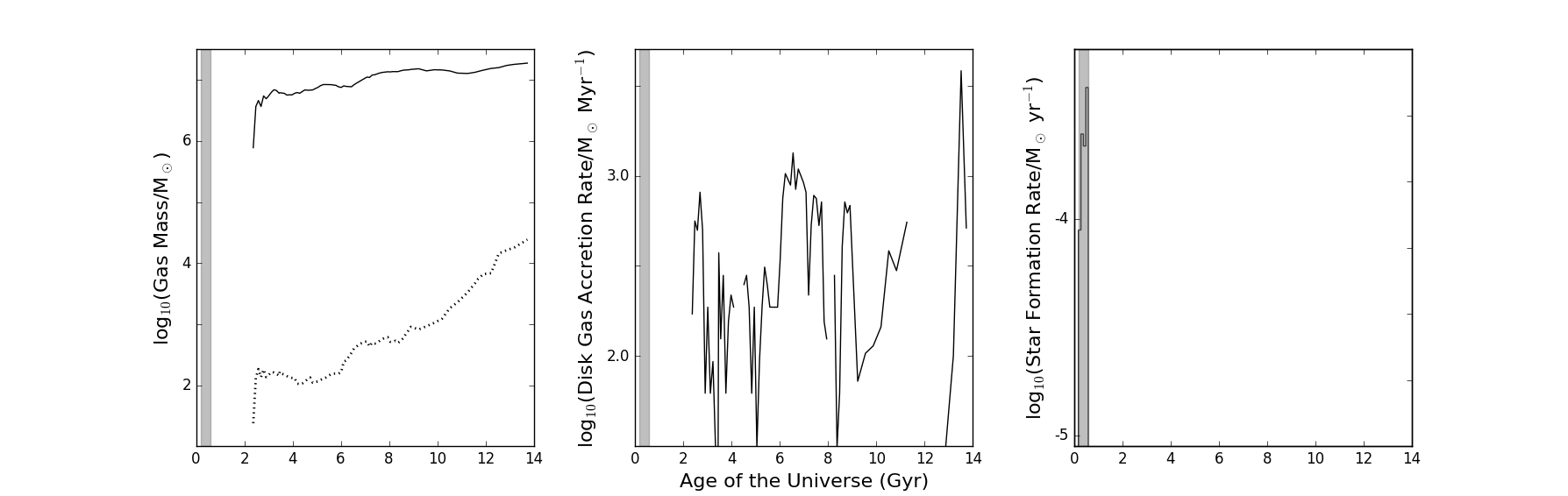}
\caption{Here, we reproduce the panels of Figure \ref{fig:story}, but for h516a, a dead galaxy. Although the galaxy never resumes star formation after being quenched by cosmic reionization, it is struck by a stream of gas from a series of nearby supernovae at t$\sim$10.5 Gyr and is building up HI when the simulation terminates. It seems likely that, were the simulation run past $z=0$, the dwarf would eventually begin forming stars again.}
\label{fig:516s}
\end{figure*}
\indent Because a few of our gappy dwarfs have a low number of star particles, we discuss resolution here.  As can be seen in Table \ref{table:prop}, three of our gappy galaxies (50\% of the sample) have only tens of star particles at $z=0$, but the other three have hundreds (or even thousands) of star particles, and would typically be considered well-resolved.  However, the small number of star particles in the three cases might lead one to question whether the gappy SFHs can be trusted. \\  
\indent In Figure \ref{fig:mgas}, we show the total gas mass in our gappy and stripped galaxies as a function of time. The main conclusion is that (until the stripped galaxies are stripped of their gas) these groups both tend to have 10$^7$ M$_\odot$ or higher gas masses, which translates into thousands of gas particles. As may be seen in Figure \ref{fig:mvir}, the gappy and stripped galaxies also have similar M$_{vir}$ (again, prior to the stripped galaxies being stripped). Thus, these two samples tend to have similar M$_{vir}$ and M$_{gas}$. The stripped galaxies also all manage to form hundreds to thousands of star particles before they are stripped of their gas. Therefore, six of the nine galaxies in the stripped+gappy samples have well-resolved star formation histories. Since these well-resolved galaxies have similar halo masses and gas masses to the three that make only tens of star particles, it indicates that the three are a special case where, indeed, some process has prevented star formation. The mass in gas is there for them to form stars if they could, but they don't. This suggests that the gaps in star formation are real. \\
\indent What might be questionable are not the gaps, but the rate of SF when it occurs in these three galaxies. In Figure \ref{fig:story}, we show the star formation rates for each of our gappy galaxies in bins of $\sim$140 Myr. All but one of our galaxies form multiple star particles in each time bin after reignition, suggesting that the SFR is well-resolved. The exception is h285a, which forms only one star particle during the final timestep of the simulation after a gap of nearly 13 Gyr. Because we cannot know if it would go on to form more stars if the simulation were run forward in time, we consider this galaxy to be a marginal case. \\
\indent We also note that one of the galaxies in the higher resolution runs, h516a, is the galaxy that appears to be on its way to becoming gappy again (see discussion of Figure \ref{fig:mag} in Section \ref{obs}). In Figure \ref{fig:516s}, we show panels for h516a similar to those for the gappy galaxies in Figure \ref{fig:story}. h516a gets compressed by gas at t$\sim$10.5 Gyr, and is building up its HI mass at $z=0$ when the simulation terminates. Prior to having its star formation shut off by reionization, this galaxy managed to form roughly 100 star particles, with a well-resolved SFH. The fact that it is on the way to being gappy indicates that gappy galaxies also occur at the higher resolutions, which we confirm in an upcoming paper (Munshi et al., in prep.). \\ \\
\bsp
\label{lastpage}
\end{document}